\documentclass[prd,aps,floats,eqsecnum,twocolumn, showpacs]{revtex4}
\usepackage{graphicx}
\usepackage{amsmath}

\newcommand{\beq}{\begin{equation}}
\newcommand{\eeq}{\end{equation}}
\newcommand{\bea}{\begin{eqnarray}}
\newcommand{\eea}{\end{eqnarray}}

\begin{document}
\title{
Probabilistic programmable quantum processors with multiple
copies of program state}
\author{Adam Brazier$^{1}$, Vladim{\'\i}r Bu\v{z}ek$^{2,3}$, and Peter L. Knight$^{1,4}$}
\address{
$^1$ Optics Section, The Blackett Laboratory, Imperial College, London SW7 2BW, United Kingdom \\
$^2$ Research Center for Quantum Information, Institute of Physics, Slovak Academy of Sciences,
D\'ubravsk\'a cesta 9,
845 11 Bratislava, Slovakia \\
$^3$ Faculty of Informatics, Masaryk University, Botanick\'a 68a,
602 00 Brno, Czech Republic \\
$^4$ National Physics Laboratory, Queen's Road, Teddington, Middleses, United Kingdom
}

\begin{abstract}We examine the execution of general U(1) transformations on programmable quantum processors. We show
that, with only the minimal assumption of availability of copies of the one-qubit program state,
that the apparent advantage of existing schemes proposed by G.Vidal {\it et al.} [Phys. Rev. Lett.
{\bf 88}, 047905 (2002)] and M.Hillery {\it et al.} [Phys. Rev. A. {\bf 65}, 022301 (2003)] to
execute a general U(1) transformation with greater probability using complex program states appears
not to hold.
\end{abstract}
\pacs{PACS Nos. 03.67.-a, 03.67.Lx}

\maketitle

\section{Introduction}
\label{4.1} In conventional classical computation, the data is manipulated by the computer (the
``processor'') according the dictates of a program. Picking the program correctly ensures that the
output data of the operation is as desired; the processor itself has general utility and can
execute many different programs.

\begin{figure}[thb]
\vspace{5mm} 
\centerline{\includegraphics[width=6cm]{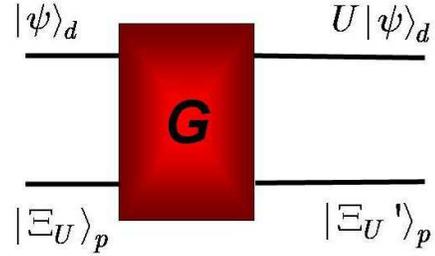}} \caption{Model of a general quantum processor.}
\label{fig1}
\end{figure}

Nielsen and Chuang \cite{Nielsen97} have investigated the possibility of a general quantum processor.
Modelling the processor as a quantum gate array into which we input a data state $| \psi
\rangle_{d}$ represented by an array of qubits, and a program state $|\Xi \rangle_{p}$ that is also
represented by an
array of qubits, we can consider the operation of the quantum processor as effected by a unitary $G$:
\beq
|\psi \rangle_{d} \otimes |\Xi \rangle_{p} \rightarrow G[|\psi\rangle_{d} \otimes |
\Xi\rangle_{p}]. \eeq In the case that the processor is to execute a particular unitary, $U$, on
the data register, we would have: \beq G[|\psi \rangle_{d} \otimes | \Xi_{U}\rangle_{p}] =
(U|d\rangle_{d}) \otimes |\Xi_{U} ^{'}\rangle_{p}, \eeq as shown in Fig.~\ref{fig1}, where $|
\Xi_{U}\rangle_{p}$  is a program state to cause the execution of $U$ on the data state and it can
be shown that the subsequent state of the program register, $|\Xi_{U} ^{\prime}\rangle_{p}$, cannot
be dependent on the data state, which for general processing will be {\em unknown} to us.
Nielsen and Chuang \cite{Nielsen97} have shown that a deterministic {\it universal} quantum
processor of  finite size does not exist.  The problem is that a new dimension must be added to
the program space for each unitary operator $U$ that one wants to be able to perform on the data
$|\psi\rangle_d$. A similar situation holds if one studies quantum circuits that implement
completely-positive, trace-preserving maps rather than just unitary operators
\cite{Hillery2001,Hillery2002a}. Some families of maps can be implemented with a finite program
space, for example, the phase damping channel, but others, such as the amplitude damping channel,
require an infinite program space. If one drops the requirement that the processor be
deterministic, then universal processors become possible
\cite{Nielsen97,Preskill1998,Vidal02,Hillery2002b}. These processors are probabilistic: they
sometimes fail, but we know when this happens.

\begin{figure}[thb]
\vspace{5mm} \centerline{\includegraphics[width=6cm]{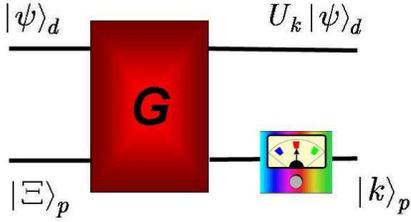}} \caption{ Model of a probabilistic
general quantum processor. On the output of the program register a measurement is performed.}
\label{fig2}
\end{figure}

In a probabilistic processor we demand that by measurement of the program register, we can tell
whether the desired unitary operation has been performed on the data state or whether some other
unitary operation has been performed upon it, i.e., that the state of the program register
associated with the execution of $U$, $| \Xi_{U}^{\prime}\rangle_{p}$, is orthogonal to the states
of the program register associated with other, undesired, outcomes on the data state (the identity
of these states of the program register will in general be dependent on the nature of the processor
itself). A model of this is shown in Fig.~\ref{fig2} where the outcome of the measurement of the
program register, $|k\rangle_{p}$, indicates which unitary operation, $U_{k}$, has been performed
on the data state.

\begin{figure}[thb]
\vspace{5mm} \centerline{\includegraphics[width=7cm]{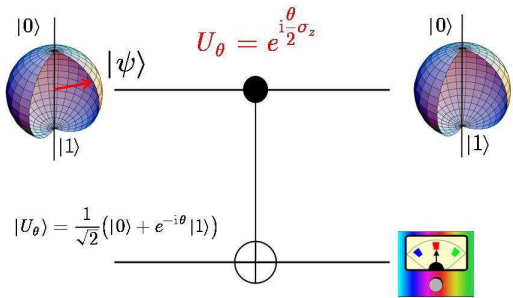}} \caption{ Model of a probabilistic
{\texttt{CNOT}} quantum processor performing the $U(1)$ rotation of the input data state
$|\psi\rangle$ by the angle $\theta$ that is encoded in the program state $|\Xi_\theta$ given by
Eq.~(\ref{eq4.1}). On the output of the program register a measurement is performed.} \label{fig3a}
\end{figure}
\begin{figure}[thb]
\vspace{5mm} \centerline{\includegraphics[width=7.3cm]{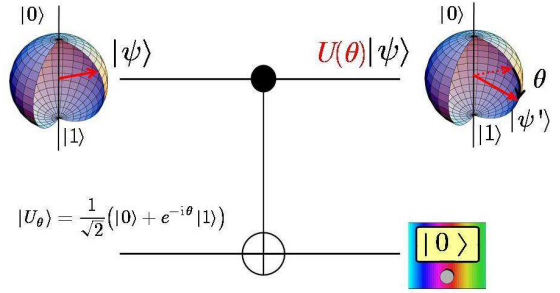}} \caption{ Model of a
probabilistic {\texttt{CNOT}} quantum processor performing the $U(1)$ rotation of the input data
state $|\psi\rangle$ by the angle $\theta$ that is encoded in the program state $|\Xi_\theta$. When
the measurement performed
 on the program
qubit result in the state $|0\rangle_p$ the desired rotation $U(\theta)$ is performed on the data
qubit. The probability of success is equal to 1/2. } \label{fig3b}
\end{figure}
\begin{figure}[thb]
\vspace{5mm} \centerline{\includegraphics[width=7.3cm]{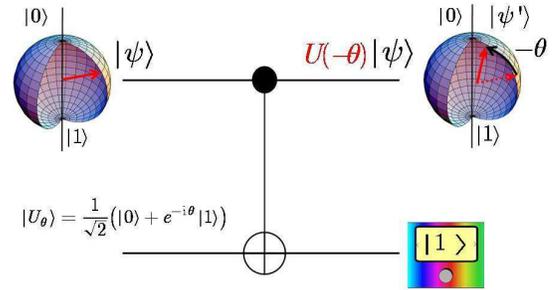}} \caption{ Model of a
probabilistic {\texttt{CNOT}} quantum processor performing the $U(1)$ rotation of the input data
state $|\psi\rangle$ by the angle $\theta$ that is encoded in the program state $|\Xi_\theta$. When
the measurement performed on the program qubit results in the state $|1\rangle_p$ the rotation
$U(-\theta)$ in the wrong direction is performed on the data qubit. The probability of this result
is equal to 1/2. } \label{fig3c}
\end{figure}

The simplest case of desired programmable operation on a qubit is the execution of a $U(1)$
transformation, $U(\theta) = e^{i\theta\sigma_{z}/2}$, upon a data qubit $|\psi\rangle_{d}=\alpha
|0\rangle_{d} +\beta|1\rangle_{d}$. Here the {\it unknown} phase of the rotation $\theta$ is
encoded in the programme state \beq |\Xi_{\theta}\rangle_{p} = \frac{1}{\sqrt{2}}\left(
|0\rangle_{p} +e^{-i\theta}|1\rangle_{p}\right)\, , \label{eq4.1} \eeq while the processor itself
is represented by a {\texttt {CNOT}} gate with data qubit as control and program qubit as target,
followed by a measurement of the program qubit in the basis $\{|0\rangle_{p},|1\rangle_{p}\}$ (see
Fig.~\ref{fig3a}).
The action of the {\texttt{CNOT}} processor on the data and the program input states is
\begin{eqnarray}
|\psi\rangle_d|\Xi_\theta\rangle_p &\longrightarrow&  \frac{1}{\sqrt{2}}
U(\theta)|\psi\rangle_d|0\rangle_p \nonumber
\\
& & + \frac{1}{\sqrt{2}} U(-\theta)|\psi\rangle_d|1\rangle_p\, .
\end{eqnarray}
From this equation we see that when a projective measurement in the computer basis
$\{|0\rangle,|1\rangle\}$  on the program qubit at the output of the {\texttt{CNOT}} is performed
and the result $|0\rangle$ is registered then the data qubit that has been prepared in an unknown
state $|\psi\rangle$ is rotated by the {\em unknown} angle $\theta$ as desired, i.e. with
probability 1/2 we obtain the state $U(\theta)|\psi\rangle_d$ (see Fig.~\ref{fig3b}).
On the other hand, when the program qubit is measured in the state $|1\rangle_p$ then the data
qubit is rotated in the opposite (``wrong'') direction, i.e. with  probability 1/2 we obtain at
the output of the probabilistic processor the state $U(-\theta)|\psi\rangle_d$ (see
Fig.~\ref{fig3c}).

In Sec.~II we shall be considering three methods of increasing the
success probability of this operation: the Vidal-Masanes-Cirac
(VMC) \cite{Vidal02} method that uses a special $N$-qubit program
state {\it iteratively} and terminates when a ``good'' result is
achieved, the Hillery-Ziman-Bu\v{z}ek (HZB) \cite{Hillery03} scheme
that uses the same program state as the VMC scheme but performs
the operation in one step, and lastly in Sec.~III we consider simply using $N$
copies of the basic program state $|\Xi_\theta\rangle$ given by
Eq.~(\ref{eq4.1}). In the latter case, we consider three
scenarios: iterative use of the program states, one-step use of
the program states and, finally, preprocessing of the program
states to produce a program state of the sort used by the VMC and
HZB schemes, which is then put through a VMC or HZB processor.
Sec.~IV is devoted to conclusions and some technical details of our
calculations are presented in Appendix.

\section{Increasing the probability of success}\label{sec.4.3}

\subsection{The VMC scheme}

The probability of successfully carrying out the $U(1)$  operation on the data qubit can be
increased through the enlargement of the program space \cite{Vidal02,Hillery03}. In the VMC scheme,
if the first operation failed, that is, we performed $U(-\theta)$ on the data state, we could
attempt to correct this by performing the rotation $U(2\theta)$ on the wrongly transformed data
state $U(-\theta)|\Psi\rangle_{d}$ and if that failed we could attempt to perform the transformation
$U(4\theta)$ on the data state $U(-3\theta)|\Psi\rangle_{d}$, etc. The $N$-qubit program state
$|\Xi_\theta^{(N)}\rangle_{\vec{p}}$ used for this iterative operation can be written as \bea
|\Xi_\theta^{(N)}\rangle_{\vec{p}} &=& |\Xi_{2^{N}\theta}\rangle_{p_1} \otimes
|\Xi_{2^{N-1}\theta}\rangle_{p_2}\otimes \ldots \otimes |\Xi_{\theta}\rangle_{p_N}
\nonumber
\\
&=& \frac{1}{\sqrt {2^{N}}}\sum_{j=0}^{2^{N}-1} e^{-ij\theta}|j\rangle_{\vec{p}},
 \label{eq4.8}
\eea with $|j\rangle_{\vec{p}} = |j_{N}\rangle_{p_N}\otimes |j_{N-1}\rangle_{p_{N-1}}\dots \otimes
|j_{1}\rangle_{p_{1}}$, where $j_{l}$ is the $l$th bit in the binary representation of $j$.

\subsection{The HZB scheme}
Instead of using iteratively the {\texttt {CNOT}} processor following Ref.~\cite{Hillery03} one can
design a general quantum processor \beq G_{dp}=\sum_{j,k=1}^{2^{N}-1}A_{jk}\otimes |j\rangle_{p} \
\ {_{p}\langle k |}\, , \label{eq4.4} \eeq where $\{j\rangle_{p}| \ j = 0,\ldots 2^{N}-1\}$ is an
orthonormal basis for the program space and the $A_{jk}$ are operators acting on the data space
such that: \beq \sum_{j=0}^{2^{N}-1} A_{xj}^{\dag}A_{jy} = \sum_{j=0}^{2^{N}-1} A_{xj}A_{jy}^{\dag}
= I_{d}\delta_{xy}. \label{eq4.5} \eeq The result of the circuit on the combined data and program
states input $|\Psi\rangle_{d}\otimes|\Xi\rangle_{p}\in {\cal H}_{d}\otimes {\cal H}_{p}$ can be
expressed as: \beq G(|\Psi\rangle_{d}\otimes |\Xi\rangle_{p}) = \sum_{j=0}^{2^{N}-1}
A_{j}(\Xi)|\Psi\rangle_{d}\otimes |j\rangle_{p}, \label{eq4.6} \eeq where the {\it program
operators } $A_{j}(\Xi)$ are given by: \beq A_{j}(\Xi) = \sum_{k=0}^{2^{N}-1}{_{p}\langle
k}|\Xi\rangle_{p}A_{jk}\, . \label{eq4.7} \eeq If the measurement of the program state returns
$|n\rangle_{p}$, then Eq.~(\ref{eq4.6}) tells us that the operation $A_{n}(\Xi)$ has been carried
out on the the data state.

To perform the $U(1)$ operation with only one iteration of the processor in the HZB scheme, we use
the same program program state as for the VMC scheme  given by Eq.~(\ref{eq4.8}). The circuit
(processor) is then determined by the operators \beq A_{jk}=\delta_{j,k}|0\rangle_{d}{_{d}\langle
0|} + \delta_{j\oplus 1,k} |1\rangle_{d}{_{d}\langle 1|} \label{eq4.9} \eeq with $\oplus$ indicating
addition modulo $2^{N}$. The program state is then measured and any result other than
$|2^{N}-1\rangle_{p}$ indicates success. The success probability for this circuit is the same as
that for the VMC circuit and it reads:
\beq
p = 1-\frac{1}{2^{N}} \label{eq4.12}\, .
\eeq
This is the
highest possible success probability achievable from the starting state
$|\Xi_\theta^{(N)}\rangle_{p}$ for a general probabilistic quantum processor \cite{Vidal02}.

\section{Using multiple copies of the basic program state}\label{sec.4.4}

\subsection{Iterative process with multiple copies of the program state $|\Xi_\theta\rangle$ \label{seciterative} }

Given that $\theta$ is not known, it is not clear how the program states for the improved schemes
above might in general be produced deterministically given no prior knowledge of $\theta$. General
execution of $U(1)$ on a data qubit using a single program qubit and a $\texttt{CNOT}$ gate is
known to be optimally achieved using the program state $|\Xi_{\theta}\rangle$ given by
Eq.~(\ref{eq4.1}) (see  Ref.~\cite{Ziman03}), so assuming the availability of this state seems a
reasonable minimal assumption. To increase the probability of success above 1/2 using just a
$\texttt{CNOT}$, we require more copies of this basic program state and, if the operation
$U(-\theta)$ has been carried out, we can reprocess the data state with a new copy of
$|\Xi_{\theta}\rangle$ and continue this process until the desired transformation has been executed
or until the available program states are exhausted \footnote{This is analogous to the Markov
process ``Gambler's ruin'', where the game is fair and the gambler has unlimited credit.}. If $N$,
the number of available copies of $|\Xi_{\theta}\rangle$, is an odd number (there is no benefit to
using an even number of program states), the probability $p$ of succeeding before running out of
copies of $|\Xi_{\theta}\rangle$ is given by the expression
\beq p=1-\frac{1}{2^{N}}\left(\begin{array}{c} N \\
(N-1)/2\end{array}\right), \label{eq4.13} \eeq and, in the limit of large $N$: \beq
p_{N\rightarrow\infty} = 1-\sqrt{\frac{2}{\pi N}}. \label{eq4.14} \eeq

\subsection{Single-shot process with multiple copies of the program state $|\Xi_\theta\rangle$\label{secsingleshot}  }

The process can be carried out with one iteration of a larger gate array where we use an odd number
of program qubits $N$ so that our combined program and data state is: \beq |\psi\rangle_{d} \otimes
|\Xi_{\theta}\rangle_{p}^{\otimes N} = \frac{|\psi\rangle_{d}}{ \sqrt{2^{N}}}\otimes
\sum_{j=0}^{2^{N}-1}e^{-i|j|\theta}|j\rangle_{\vec{p}}, \label{eq4.15} \eeq where $|j|$ is the
Hamming weight of the binary representation of $j$ and we use the same basis for the program space
as previously. Putting $A_{kk} = |0\rangle_{d}{_d}\langle 0|$ as before, we select the position of
the terms $A_{jk}= |1\rangle_{d}{_d}\langle 1|$ according to the Hamming weight of the $j$ and $k$
such that \beq |k| = |j| +1 \label{eq4.16} \eeq to the largest extent possible so that
Eq.~(\ref{eq4.5}) is obeyed and we can position the other terms arbitrarily so as to respect
Eq.~(\ref{eq4.5}). Where we can give the $A_{jk}$ values according to Eq.~(\ref{eq4.16}),
measurement in the program basis will, up to global phase, ensure that the data qubit has been
transformed by $U(\theta)$. The rows (values of $j$) where $A_{jk}=|1\rangle_{d}{_d}\langle1|$ are
not positioned according to $|k|=|j|+1$ indicate measurement outcomes where the desired
transformation has not been carried out but instead a rotation through some negative multiple of
$\theta$ has occurred. The number $R$ of rows that cannot be created so that Eq.~(\ref{eq4.16}) is
obeyed is given by: \beq R = \left(\begin{array}{c} N\\ (N-1)/2\end{array}\right) \label{eq4.17}
\eeq Each (incorrect) program operator corresponding to one of these rows has probability $2^{-N}$
so again the success probability is given by Eq.~(\ref{eq4.13})\footnote{In this case, unlike the
VMC and HZB schemes, the distribution of particular incorrect results can differ according to how
the $A_{jk}$ are selected, although the overall probability of success is unchanged.}.

\subsection{Preprocessing}\label{sec4.5}

If we wish to use, from a starting state of multiple copies of $|\Xi_\theta\rangle$, the VMC or HZB
schemes, we can process these copies to produce a state of the form given in Eq.~(\ref{eq4.8}) that
can then be used as the program state for the VMC or HZB processors. The $X-$qubit program state
$|\Xi_\theta^{(X)}\rangle_{p}$ can be probabilistically constructed from a minimum of $N=2^{X}-1$
copies of $|\Xi_\theta\rangle$, and so it is possible, by preprocessing these copies of
$|\Xi_\theta\rangle$, to construct, with some probability, a state $|\Xi(\theta)_{s}\rangle_{p}$
where $s \le X$. A preprocessing scheme that produces the same overall probability of success, in
executing $U(\theta)$ on a data qubit, as the schemes in Sec.~\ref{sec.4.3} can be constructed by
permuting the phases in $|\Xi_\theta\rangle^{\otimes 2^{X}-1}$ and making a measurement in the
computational basis, initially on $2^{X}-1-X=M$ of the qubits.

We give two specific examples, of preprocessing. Firstly, we will assume to have three identical
program states  $|\Xi_\theta\rangle^{\otimes 3}$. Then we will consider the case with seven identical
program states, i.e. $|\Xi_\theta\rangle^{\otimes 7}$. Using 3 and 7 program state we can probabilistically
prepare the program states $|\Xi_\theta^{(2)}\rangle_{p}$ and $|\Xi_\theta^{(3)}\rangle_{p}$, respectively.
In the Appendix we will quote the result for general $N$.

\subsubsection{Preprocessing with three copies of $|\Xi_\theta\rangle$}

We have that:


\bea |\Xi_{\theta}\rangle^{\otimes 3} &=& \frac{1}{2\sqrt{2}}(|000\rangle + e^{-i\theta}|001\rangle
+ e^{-i\theta}|010\rangle\nonumber\\ &+& e^{-2i\theta}|011\rangle + e^{-i\theta}|100\rangle +
e^{-2i\theta}|101\rangle \nonumber\\&+& e^{-2i\theta}|110\rangle + e^{-3i\theta}|111\rangle),
\label{eq:3 qubit product state}\eea in the computational basis. The states that can be constructed
from this are $|\Xi^{(1)}_{\theta}\rangle$ and $|\Xi^{(2)}_{\theta}\rangle$ which are, up to global
phase and in the computational basis:

\beq |\Xi^{(1)}_{\theta}\rangle = \frac{1}{\sqrt{2}}(|0\rangle) + e^{-i\theta}|1\rangle)
\label{eq:one qubit Xi defined } \eeq and

\bea |\Xi^{(2)}_{\theta}\rangle &=&  |\Xi_{2\theta}\rangle \otimes |\Xi_{\theta}\rangle \nonumber\\
&=& \frac{1}{2} (|00\rangle) + e^{-i\theta}|01\rangle + e^{-i2\theta}|10\rangle +
e^{-3i\theta}|11\rangle ). \nonumber \\\label{eq: 2 qubit Xi defined}\eea The state is permuted,
which has the effect of reassigning the phases:

\bea
\nonumber
|\Xi_{\theta}\rangle^{\otimes 3} &\mapsto& \frac{1}{2\sqrt{2}}(|000\rangle +
e^{-i\theta}|001\rangle + e^{-2i\theta}|010\rangle \nonumber \\&+& e^{-3i\theta}|011\rangle  +
e^{-i\theta}|100\rangle + e^{-2i\theta}|101\rangle \nonumber\\&+& e^{-i\theta}|110\rangle +
e^{-2i\theta}|111\rangle)\\
&=&\left(\frac{|0\rangle}{\sqrt{2}}\otimes |\Xi^{(2)}_{\theta}\rangle\right)\nonumber\\ &+&
\left(\frac{e^{-i\theta}|1\rangle}{\sqrt{2}} \otimes \left(\frac{|0\rangle}{\sqrt{2}} \otimes
|\Xi^{(1)}_{\theta}\rangle + \frac{|1\rangle}{\sqrt{2}}\otimes |\Xi^{(1)}_{\theta}\rangle \right)
\right).\nonumber\\\label{eq:3 qubit end state}
\eea
Eq.~(\ref{eq:3 qubit end state}) shows that a measurement on the first (leftmost in the
right-hand-side of the previous equation) qubit would either
give $|\Xi^{(2)}_{\theta}\rangle$ upon measurement outcome $|0\rangle$, or a state, on measurement
outcome $|1\rangle$ which can be reduced to $|\Xi^{(1)}_{\theta}\rangle$, up to global phase, by
measurement of the remaining leftmost qubit. Each of these final results occurs with probability
$1/2$ and so, using Eq.~(\ref{eq4.12}), we find that the overall probability of successfully
executing the operation $U(\theta)$ following preprocessing of the state and then input of the
outcome, as a program state, into a HZB or VMC process is $5/8$, which is in fact the same as that
for iterative or single-shot processing of the state $|\Xi_\theta\rangle^{\otimes 3}$ from Secs.
\ref{seciterative} and \ref{secsingleshot}, as can be calculated from Eq.~(\ref{eq4.13}).

The preprocessing transformation (\ref{eq:3 qubit end state}) can be easily realized using a single CNOT gate with
the second qubit in Eq.(\ref{eq:3 qubit product state})
playing the role of a control with the first qubit acting as a target.

\subsubsection{Preprocessing with 7 copies of the program state $|\Xi_\theta\rangle$}

In considering the preprocessing of $|\Xi_\theta\rangle^{\otimes 7}$ we introduce a technique for
permutation design that is helpful in describing the derivation of the general preprocessing
procedure for $|\Xi_\theta\rangle^{\otimes N}$.

The starting point is the state:

\bea |\Xi_{\theta}\rangle^{\otimes 7} &=& \frac{1}{\sqrt{128}}\sum_{j=0} ^{127} e^{-i|j|}|j\rangle
\label{eq:7 qubit product state}
\\&=& \frac{1}{\sqrt{128}}\sum_{p=0} ^{15}|p\rangle \otimes \sum_{q=0}
^{7}e^{-i(|q|+|p|)\theta}|q\rangle \nonumber\eea
and the procedure is to
perform a permutation of the state so that measurement of the first four qubits in the
computational basis will yield either $|\Xi^{(3)}_{\theta}\rangle$ or a state from which
measurement of the one or two remaining leftmost qubits will yield $|\Xi^{(2)}_{\theta}\rangle$ or
$|\Xi^{(1)}_{\theta}\rangle$, respectively, up to a global phase. The numbers of terms with each
phase are given by

\beq \nonumber
\begin{array}{|c||c|c|c|c|c|c|c|c|}\hline
-ik\theta & 0 & -i\theta & -2i\theta& -3i\theta& -4i\theta& -5i\theta& -6i\theta & -7i\theta
\\\hline
m & 1 & 7 & 21 & 35 & 35 & 21 & 7 & 1\\
\hline
\end{array}\eeq and the aim is to allocate those phases to terms so that, upon measurement of the
leftmost four qubits, the state is either projected into $|\Xi^{(3)}_{\theta}\rangle$ or else a
state from which further measurement will project into $|\Xi^{(2)}_{\theta}\rangle$ or
$|\Xi^{(1)}_{\theta}\rangle$ up to global phase. Noting that one set of the phases $0$, $-i\theta$,
$-2i\theta$, $-3i\theta$, $-4i\theta$, $-5i\theta$, $-6i\theta$, $-7i\theta$ are available, the
permutation can be constructed so that the 4-qubit measurement outcome $|0\rangle$ in
Eq.~(\ref{eq:7 qubit product state}) is:

\bea  \frac{1}{4}|0\rangle &\otimes& \frac{1}{\sqrt{8}}\left(|0\rangle + e^{-i\theta}|1\rangle +
e^{-2i\theta}|2\rangle + e^{-3i\theta}|3\rangle \right. \nonumber\\
 &+& \left.e^{-4i\theta}|4\rangle + e^{-5i\theta}|5\rangle + e^{-6i\theta}|6\rangle
 e^{-7i\theta}|7\rangle \right)\nonumber \\&=& \frac{1}{4}|0\rangle \otimes
|\Xi^{(3)}_{\theta}\rangle \, . \eea
The following phases
\beq \nonumber
\begin{array}{|c||c|c|c|c|c|c|c|c|}\hline
-ik\theta & 0 & -i\theta & -2i\theta& -3i\theta& -4i\theta& -5i\theta& -6i\theta & -7i\theta
\\\hline
m & 0 & 6 & 20 & 34 & 34 & 20 & 6 & 0\\
\hline
\end{array}\eeq remain unassigned in the permutation.
It can be seen that the terms associated with the 4-qubit measurement outcome $|1\rangle$
cannot constitute $|\Xi^{(3)}_{\theta}\rangle$, as the requisite phases have already been allocated
to the terms associated with the measurement outcome $|0\rangle$. However, allocation of the
phases $-i\theta$, $-2i\theta$, $-3i\theta$ and $-4i\theta$ and also $-3i\theta$, $-4i\theta$,
$-5i\theta$ and $-6i\theta$ allows that the permutation can be designed such that the 4-qubit
measurement outcome $|1\rangle$ is
\bea
\frac{1}{4} |1\rangle &\otimes& \frac{1}{\sqrt{8}}\left( e^{-i\theta}|0\rangle +
e^{-2i\theta}|1\rangle + e^{-3i\theta}|2\rangle + e^{-4i\theta}|3\rangle\right. \nonumber\\ &+&
\left. e^{-3i\theta}|4\rangle + e^{-4i\theta}|5\rangle + e^{-5i\theta}|6\rangle + e^{-6i\theta}|7\rangle  \right) \nonumber\\
&=& \frac{1}{4} |1\rangle \otimes \frac{1}{\sqrt{2}}\left(|0\rangle \otimes
e^{-i\theta}|\Xi^{(2)}_{\theta}\rangle + |1\rangle \otimes e^{-3i\theta}|\Xi^{(2)}_{\theta}\rangle
\right)\, .
\nonumber \\ \eea
A further measurement of the leftmost remaining qubit will project the state of remaining qubits
into $|\Xi^{(2)}_{\theta}\rangle$ up to a global phase of $e^{-i\theta}$ or $e^{-3i\theta}$. The
remaining phases are

\beq \nonumber
\begin{array}{|c||c|c|c|c|c|c|c|c|}\hline
-ik\theta & 0 & -i\theta & -2i\theta& -3i\theta& -4i\theta& -5i\theta& -6i\theta & -7i\theta
\\\hline
m & 0 & 5 & 19 & 32 & 32 & 19 & 5 & 0\\
\hline
\, .\end{array}\eeq The same allocation can be performed for the 4-qubit measurement outcomes
$|2\rangle$ to $|6\rangle$. The remaining unallocated phases are

\beq \nonumber
\begin{array}{|c||c|c|c|c|c|c|c|c|}\hline
-ik\theta & 0 & -i\theta & -2i\theta& -3i\theta& -4i\theta& -5i\theta& -6i\theta & -7i\theta
\\\hline
m & 0 & 0 & 14 & 22 & 22 & 14 & 0 & 0\\
\hline
\end{array}
\eeq and it is therefore possible to construct the permutation so that the measurement outcomes
$|7\rangle$ to $|13\rangle$ are
\bea  \frac{1}{4} |j\rangle &\otimes& \frac{1}{\sqrt{8}}\left( e^{-2i\theta}|0\rangle +
e^{-3i\theta}|1\rangle + e^{-4i\theta}|2\rangle + e^{-5i\theta}|3\rangle \right. \nonumber\\ &+&
\left. e^{-2i\theta}|4\rangle + e^{-3i\theta}|5\rangle + e^{-4i\theta}|6\rangle + e^{-5i\theta}|7\rangle  \right) \nonumber\\
&=& \frac{1}{4} |j\rangle \otimes \left( \frac{|0\rangle + |1\rangle}{\sqrt{2}}\right)\otimes
e^{-2i\theta}|\Xi^{(2)}_{\theta}\rangle \, ; \ \ j=7 \dots 13\, . \nonumber \\ \eea
Any measurement on
the leftmost remaining qubit projects into the state $e^{-2i\theta}|\Xi^{(2)}_{\theta}\rangle$.
Finally, the remaining phases,

\beq \nonumber
\begin{array}{|c||c|c|c|c|c|c|c|c|}\hline
-ik\theta & 0 & -i\theta & -2i\theta& -3i\theta& -4i\theta& -5i\theta& -6i\theta & -7i\theta
\\\hline
m & 0 & 0 & 0 & 8 & 8 & 0 & 0 & 0\\
\hline
\end{array}
\eeq are allocated to the 4-qubit measurement outcomes $|14\rangle$ and $|15\rangle$ like so

\bea \frac{1}{4} |l\rangle &\otimes& \frac{1}{\sqrt{8}}\left( e^{-3i\theta}|0\rangle +
e^{-4i\theta}|1\rangle + e^{-3i\theta}|2\rangle + e^{-4i\theta}|3\rangle \right. \nonumber \\ &+&
\left. e^{-3i\theta}|4\rangle + e^{-4i\theta}|5\rangle + e^{-3i\theta}|6\rangle + e^{-4i\theta}|7\rangle  \right) \nonumber\\
&=& \frac{1}{2} |14\rangle \otimes \left(\frac{|0\rangle + |1\rangle + |2\rangle + |3\rangle}{2}
\right) \otimes e^{-3i\theta}|\Xi^{(1)}_{\theta}\rangle\nonumber \\\eea
with  $l=14,15$. A
measurement of the two leftmost remaining qubits will project the remaining qubits into the state
$e^{-3i\theta}|\Xi^{(1)}_{\theta}\rangle$. Thus, the permutation construction is complete and the
overall, permuted state, $|\tilde{\Xi}_{\theta}\rangle_{7}$ is given by:
\bea |\tilde{\Xi}_{\theta}\rangle_{N} = \frac{1}{4}|0\rangle &\otimes& |\Xi^{(3)}_{\theta}\rangle \nonumber \\
+ \frac{1}{4} \sum_{k=1} ^{7}|k\rangle &\otimes& \frac{1}{\sqrt{2}}\left(|0\rangle \otimes
e^{-i\theta}|\Xi^{(2)}_{\theta}\rangle + |1\rangle \otimes e^{-3i\theta}|\Xi^{(2)}_{\theta}\rangle
\right)\nonumber \\ + \frac{1}{4} \sum_{k=8} ^{13}|k\rangle &\otimes& \left( \frac{|0\rangle +
|1\rangle}{\sqrt{2}}\right)\otimes e^{-2i\theta}|\Xi^{(2)}_{\theta}\rangle \nonumber \\
+\frac{1}{2} \sum_{k=14}^{15}|k\rangle &\otimes& \left(\frac{|0\rangle + |1\rangle + |2\rangle +
|3\rangle}{2} \right) \otimes e^{-3i\theta}|\Xi^{(1)}_{\theta}\rangle. \nonumber \\\eea The
probability of the preprocessing procedure, following the 4-qubit measurement in the computational
basis, producing the outcome  $|\Xi^{(3)}_{\theta}\rangle$ is $1/16$, that of producing outcome
$|\Xi^{(2)}_{\theta}\rangle$ is $13/16$ and that of producing outcome $|\Xi^{(1)}_{\theta}\rangle$
is $1/8$. The overall probability, p, then, of achieving the rotation $U(\theta)$ from the starting
state $|\Xi_\theta\rangle^{\otimes 7}$ by preprocessing and then input of the preprocessed state
into the VMC or HZB processors, is
\beq p =\left(\frac{7}{8}\times\frac{1}{16}\right) + \left(\frac{3}{4}\times\frac{13}{16}\right) +
\left(\frac{1}{2}\times\frac{1}{8}\right) = \frac{93}{128},\eeq which is the same as the iterative
or single-shot procedures outlined in Secs. \ref{seciterative} and \ref{secsingleshot}, as can
be confirmed with use of Eq.~(\ref{eq4.13}). It should be noted that the permutation outlined
above is not unique and that other permutations could be devised to achieve the same overall
success probability.

\subsubsection{Preprocessing with $N$ copies of the program state $|\Xi_\theta\rangle$}

The equivalence of the iterative, single-shot and preprocessing schemes can be shown to be true in
general for states of $N=2^{X}-1$, $X=1,2, \ldots$ copies of $|\Xi_\theta\rangle$, as described in
Appendix, so that the overall success probability from a preprocessing of the state
$|\Xi_\theta\rangle^{\otimes N}$ as described above, followed by input of the result of the
preprocessing into a VMC or HZB processor, is the same as that in Eq.~(\ref{eq4.13}), i.e., \bea
p=1-\frac{1}{2^{N}}\left(\begin{array}{c}N \\ (N-1)/2\end{array}\right),\eea and thus we see that
the use of the VMC or HZB schemes holds no advantage in terms of overall success probability when we
are constrained to start with $|\Xi_\theta\rangle^{\otimes N}$. This is the main result of our
paper.

\section{Conclusion}

If we have no reason to assume that previous operations have produced a program state
$|\Xi_\theta^{(N)}\rangle_{\vec{p}}$, then it is reasonable to assume that we only have access to
copies of the basic program state $|\Xi_{\theta}\rangle$; in this case there is no advantage, in
terms of probability of success, in using the more sophisticated VMC and HZB schemes to execute the
desired $U(1)$ operation because what we gain from those schemes we lose in producing the correct
input program state. It appears that all strategies, in practice, give the same probability of
success in executing the desired $U(1)$ rotation on a qubit. There may, however, be contextual
advantages to the preprocessing scheme, for example, if the program state is to be teleported to a
remote location before execution of the program; in this case, preprocessing means that the number
of qubits to be transported is significantly lessened, which would be helpful if teleportation
resources are scarce. On the other hand, if teleportation is unreliable but teleportation resources
are not scarce, it might be better to teleport the copies of the basic program state as is, because
the effect of losing a program qubit is not so great as in the case of sending the preprocessed
states.

It is an open question as to whether a similar situation holds for the execution of the most
general unitary operations on a qubit, the $SU(2)$ operations (see, for example,
Refs.~\cite{Nielsen97, Hillery03}).

\noindent {\bf Acknowledgements}\newline We thank Mario Ziman for useful discussions. This work was
supported in part by  the European Union projects QGATES  and CONQUEST, and by the UK
Engineering and Physical Sciences Research Council.

\appendix
\section{PREPROCESSING SUCCESS PROBABILITY}\label{app4.1}

We have seen how the preprocessing scheme works for $|\Xi_{\theta}\rangle^{\otimes 3}$ and
$|\Xi_{\theta}\rangle^{\otimes 7}$, and that it produces the same probability for success as the
one-shot and iterative schemes with the same starting states. The general scheme for preprocessing
$2^{X}-1$ copies of the basic program state, where $X$ is an integer, is an extension of the method
used in Sec.~\ref{sec4.5}. Given $|\Xi_{\theta}\rangle^{\otimes 2^{X}-1}$, the best VMC/HZB program
state that can be produced is $|\Xi^{(X)}_{\theta}\rangle$, because the phases start at $0$, rise in
increments of $-i\theta$ and the largest phase in $|\Xi_{\theta}\rangle^{\otimes 2^{X}-1}$ is
$-i(2^{X}-1)\theta$, which is also the biggest phase in $|\Xi^{(X)}_{\theta}\rangle$, where the
phases also rise in increments of $-i\theta$ from a phase of $0$. The strategy will be to permute
the phases on the $2^{2^{X}-1}$ terms in $|\Xi_{\theta}\rangle^{\otimes 2^{X}-1}$, where the number
of terms with each phase is binomially distributed, in a useful way and then measure the leftmost
$M=2^{X}-1-X$ qubits to project into a remainder $X$-qubit state which will be
$|\Xi^{(X)}_{\theta}\rangle$ or some other state which, upon further measurements of leftmost
remaining qubits, will be projected into $|\Xi^{(r)}_{\theta}\rangle$ where $r \in \{1, 2, \ldots
X-1 \}$, up to a global phase, as was the case in the examples in Sec.~\ref{sec4.5} for $X=2$ and
$X=3$, i.e., the permutation achieves:

\bea \frac{1}{\sqrt{2^{2^{X}-1}}}\sum_{j=0}^{2^{2X-1}} e^{i|j|\theta}|j\rangle \rightarrow
\frac{1}{\sqrt{2^{M}}} \sum_{k=0}^{2^{M}-1}|k\rangle\otimes|k_{\Xi}\rangle\, .\eea The $X$-qubit
states $|k_{\Xi}\rangle$ are given by \bea |k_{\Xi}\rangle =
\sum_{l=1}^{X}\sum_{t=0}^{2^{X-l}-1}a_{lt}^{k}\left(|t\rangle \otimes |\Xi_{\theta}^{(l)}\rangle
\right),\eea where the $|t\rangle$ are $(X-l)$-qubit computational basis states and normalisation
requires that
\beq \sum_{l=1}^{X}\sum_{t=0}^{2^{X-l}-1}|a_{lt}^{k}|^{2} = 1 \eeq and we note that not all of the
$a_{l}^{k}$ need be non-zero. In addition,  these coefficients
have to be such that the measurement outcomes subsequent
to the initial M-qubit measurement are entangled with a particular eventual outcome,
i.e., one of the $|\Xi^{l}\rangle$ so that if we measure the initial $M$ qubits then carry out some
more measurements, that the final measurement outcome $|t\rangle$ tells us what VMC/HZB program state we have.

The allocation of phases in the construction of the permutation is done in the same way as was
shown in some detail for $|\Xi_{\theta}\rangle^{\otimes 7}$, which is to say, first one of each
phase is allocated to the $2^{X}$ terms that will produce $|\Xi^{(X)}_{\theta}\rangle$ upon one
outcome of the measurement of the $M$ leftmost qubits. Following that, phases $-i\theta \ldots
-2^{X-1}i\theta$ and $-i(2^{X-1}-1)\theta \ldots -i(2^{X}-2)\theta$ (that was $-i\theta$ to
$-4i\theta$ and $-3i\theta$ to $-6i\theta$ in the $X=3$, $N=7$ examples) are allocated to sets of
$2^{X-1}$ terms until the phases $-i\theta$ and $-i(2^{X}-2)\theta$ are exhausted and then phases
$-2i\theta \ldots -(2^{X-1}+1)i\theta$ and $-(2^{X-1}-2)i\theta \ldots -i(2^{X}-3)$ are allocated,
etc, until there are only $2^{X-1}-2$ different phases left available (the ``middle'' $2^{X-1}-2$
phases if laid out as in the tables of Sec.~\ref{sec4.5}). These groups of terms will be those that
realise $|\Xi^{(X-1)}_{\theta}\rangle$ post-measurement. Following this, the procedure is to
allocate groups of $2^{X-2}$ phases so as to create groups of terms that will realise
$|\Xi^{(X-2)}_{\theta}\rangle$ post-measurements, and so on, until the last remaining phases,
$-i(2^{X-1}-1)\theta$ and $-2^{X-1}i\theta$, are allocated to the terms that will produce
$|\Xi^{(1)}_{\theta}\rangle$ post-measurements.

The key facts here are that all of the phases can be allocated in this way to a group of terms
associated, post-measurements, with the realisation of a state $|\Xi(\theta)_{s}\rangle_{p}$ where
$s \le X$, as a little thought will show. Furthermore, with the phases allocated in this way, every
group of phases allocated contains the ``middle'' two phases, $-i(2^{X-1}-1)\theta$ and
$-2^{X-1}i\theta$. Thus, the number of groups of phases, $W$, is equal to the number of terms in
$|\Xi_{\theta}\rangle^{\otimes 2^{X}-1}$ that have phase $-i(2^{X-1}-1)\theta$ or
$-2^{X-1}i\theta$, i.e.,

\beq W = \left(\begin{array}{c} 2^{X}-1\\
(2^{X}-2)/2\end{array}\right).\eeq If the number of groups corresponding to
$|\Xi(\theta)_{s}\rangle_{p}$ is $W_{s}$, then, because each individual phase from the terms in
$|\Xi_{\theta}\rangle^{\otimes 2^{X}-1}$ is allocated to one of these groups,

\bea \sum_{s=1} ^{X}W_{s}
= W=\left(\begin{array}{c} 2^{X}-1\\
(2^{X}-2)/2\end{array}\right). \label{eq:number of middle terms}\eea Additionally, because all of
the $2^{2^{X}-1}$ terms in $|\Xi_{\theta}\rangle^{\otimes 2^{X}-1}$ end up permuted into one of
these sets, and because each set of form $|\Xi^{(s)}_{\theta}\rangle$ contains $2^{s}$ terms, with
$W_{s}$ sets of form $|\Xi^{(s)}_{\theta}\rangle$ and $s$ different types of set, then

\beq \sum_{s=1}^{X}2^{s}W_{s} = 2^{(2^{X}-1)} \label{eq:total number of terms}. \eeq The
probability, $q_{s}$, that the final result is $|\Xi^{(s)}_{\theta}\rangle$ following
measurement(s), can be expressed in terms of $W_{s}$. It is equal to the number of terms that
belong in sets of form $|\Xi^{(s)}_{\theta}\rangle$ divided by the total number of terms, ie:

\beq q_{s} = \frac{2^{s}W_{s}}{2^{(2^{X}-1)}}.\eeq

Each state $|\Xi^{(s)}_{\theta}\rangle$ will, if it is the outcome of the calculation, succeed in
the VMC/HZB scheme with a probability $p_{s}$ given by:

\beq p_{s} = 1 - \frac{1}{2^{s}} \eeq from Eq.~(\ref{eq4.12}).

The total success probability, $p_{X}$, from preprocessing $|\Xi_{\theta}\rangle^{\otimes 2^{X}-1}$
followed by the input of the resulting state as the program state into the HZB/VMC scheme, is

\bea p_{X} &=& \sum_{s} ^{X} p_{s} q_{s} \nonumber \\
&=& \frac{1}{2^{(2^{X}-1)}}\left(\sum_{s=1}^{X}2^{s}W_{s} - \sum_{s=1}^{X}W_{s}\right) \nonumber
\\
&=& 1 - \frac{1}{2^{(2^{X}-1)}}\left(\begin{array}{c} 2^{X}-1\\
(2^{X}-2)/2\end{array}\right). \label{eq:overall succ prob in terms of X}\eea where the last step
was achieved using Eqs.~(\ref{eq:number of middle terms}) and (\ref{eq:total number of terms}).
The total number of basic program qubits, $N$, is given by:

\beq N = 2^{X}-1 \eeq and substituting this into Eq.~(\ref{eq:overall succ prob in terms of X}),
the overall probability of success, $p$, is given by:

\bea p=1-\frac{1}{2^{N}}\left(\begin{array}{c}N \\ (N-1)/2\end{array}\right).\eea This is the same
result as for the single-shot and iterative procedures on $|\Xi_{\theta}\rangle^{\otimes N}$ and so
preprocessing gives the same overall probability of success as in those case and the result is
proved. Although this calculation is based on a specific method of allocation of the states, it
will be true for any permutation allocation that puts all of the phases in the state
$|\Xi_\theta\rangle^{\otimes 2^{X}-1}$ into a grouping that produces a state
$|\Xi(\theta)_{s}\rangle_{p}$, $s \le X$ and in which each grouping contains the two ``middle''
phases, i.e., the phases $-i(2^{X-1}-1)\theta$ and $-2^{X-1}i\theta$.

\end{document}